\documentstyle[12pt]{article}
\begin{document}

\begin{center}
\begin{large}
{\bf Statistics transmutations in two-dimensional systems\\
and the fractional quantum Hall effect}
\end{large}
\end{center}

\vspace{1cm}

\noindent Piotr Sitko, Institute of Physics,
Technical University of Wroc\l{}aw,\\ Wyb. Wyspia\'{n}skiego
 27, 50-370 Wroc\l{}aw, Poland.

 \vspace{1cm}
\begin{center}

Abstract

\end{center}
Statistics transmutations to composite fermions in
fractional-quantum-Hall-effect systems
are considered
in the Hartree-Fock approximation and in the RPA.
The Hartree-Fock ground-state energy shows that the transmutations
 are not energetically preferable.
Within the RPA it is found that the system exhibits a fractional
quantum Hall effect.

\vspace{2cm}
\newpage
\noindent{\bf 1.  Introduction}

The fractional quantum Hall effect (FQHE)  brings
many intrinsic questions to physics
of two-dimensional systems. Owing to the idea of
Laughlin \cite{Laughlin}  one can explain experimentally observed
phenomena, however, an origin of the Laughlin wave function still
remains somewhat mysterious.

An alternative approach to understanding the nature of the FQHE was
proposed by Jain \cite{Jain,Greiter} within the idea of
transmutability of statistics in two-dimensional systems.
Jain noted that in 2D the antisymmetrity of the many-particle wave
function  is not an unambiguous criterion of quantum statistics.
The antisymmetrity is held by a class of quantum-statistics particles called
composite fermions \cite{Jain}. An interchange of two
composite fermions produces a
phase factor of $e^{i(2p+1)\pi}$ (p - an integer number). Note that
this is also the case of the Laughlin wave function.

Statistics transmutations to composite fermions are described by a Chern-Simons
theory with an even number of flux quanta attached to each
electron \cite{Lopez1,Lopez2}. In the mean field approximation
Chern-Simons interactions produce the average statistical field\\
$B^{s}=-2p\frac{hc}{e}\rho$ which partially cancels the external
magnetic field. Thus, the effective field acting on electrons is
$B^{ef}=B^{ex}+B^{s}$. We predict an analog of the quantum Hall
effect when, effectively, $n$ Landau levels a completely filled, i.e.
$B^{ef}=\frac{1}{n} \frac{hc}{e}\rho$. Hence,
$B^{ex}=\frac{2pn+1}{n} \frac{hc}{e}\rho$, which means that from the
point of view of the external magnetic field the lowest Landau level
is filled in the fraction $\nu=\frac{n}{2pn+1}$. We do not consider
the case of $|B^{s}|>B^{ex}$ when $B^{ef}\uparrow\downarrow B^{ex}$
and one expects an opposite Hall effect.

The aim of this paper is to find, within the Hartree-Fock approximation (HFA),
the ground state and the ground-state energy of the system. We obtain
also the RPA response function, and thus collective modes and a Hall
conductance.

\noindent{\bf 2. Hartree-Fock ground state}

Let us start from the Hamiltonian of anyons (in the fermion
representation) in the external magnetic
field  \cite{Chen}:
\begin{equation}
H=\frac{1}{2m}\sum_{i=1}^{N}({\bf P}_{i}+\frac{e}{c}{\bf
A}_{i}+\frac{e}{c}{\bf
A}_{i}^{ex})^{2}
\end{equation}
where	Chern-Simons interactions are given by
\begin{equation}
{\bf A}_{i}=\frac{-2p \hbar c}{e}	\hat{\bf z}\times \sum_{j\neq i}
\frac{({\bf r}_{i}-{\bf r}_{j})}{|{\bf r}_{i}-{\bf r}_{j}|^{2}}\;\; ,
\end{equation}
$\hat{\bf z}$ - a perpendicular to the plane unit vector.
The average statistical  field vector potential is found via the equation:
\begin{equation}
\bar{\bf A}_{i}=\frac{-2p \hbar c \rho}{e}\hat{\bf z}\times\int
\frac{({\bf r}_{i}-{\bf r})}{|{\bf r}_{i}-{\bf r}|^{2}}
d^{2}{\bf r}=\frac{1}{2}B^{s}\hat{\bf z}\times {\bf r}_{i} \; .
\end{equation}
The Hamiltonian (1) for  integer values of $p$
describes the statistics transmutations to composite fermions (in
the system of electrons in the external magnetic field) and for
other values of $p$ - the transmutations to anyons with  statistics
parameters
$\Theta = (1+2p)\pi$ (an exchange of two anyons produces a phase
factor of $e^{i\Theta}$).
In this paper we assume that $B^{ef}=\nabla\times {\bf
A}^{ef}=B^{ex}+B^{s}=\frac{1}{n} \frac{hc}{e}\rho$ and in the ground
state one has $n$ completely filled Landau levels.

Dividing the Hamiltonian (1) into one-, two- and three-body parts one
finds the Hartree-Fock equations of the ground-state energy
 \cite{Hanna,Fetter,Sitko}:
\begin{equation}
<H_{a}>=\frac{1}{2m}\sum_{l}\int d1 \phi_{l}^{+}(1)({\bf
P}_{1}+\frac{e}{c}{\bf A}^{ef})^{2}\phi_{l}(1)\;\; ,
\end{equation}
\begin{equation}
<H_{b}>=-\frac{e}{mc}\sum_{l,m} \int d1d2\phi_{l}^{+}(1)
\phi_{m}^{+}(2){\bf A}_{12}({\bf P}_{1}+\frac{e}{c}{\bf A}^{ef})
\phi_{l} (2)
\phi_{m} (1)\; ,
\end{equation}
\begin{equation}
<H_{c}>=\frac{e^{2}}{2c^{2}m}\sum_{l,m} \int d1d2\phi_{l}^{+}(1)
\phi_{m}^{+}(2)|{\bf A}_{12}|^{2}\phi_{l}(1)
\phi_{m}(2)\; ,
\end{equation}
\begin{equation}
<H_{d}>=-\frac{e^{2}}{2c^{2}m}\sum_{l,m}\int d1d2\phi_{l}^{+}(1)
\phi_{m}^{+}(2)|{\bf A}_{12}|^{2}\phi_{l}(2)
\phi_{m}(1)\; ,
\end{equation}
\begin{equation}
<H_{e}>=\frac{e^{2}}{c^{2}m}\sum_{l,m,k}\int
d1d2d3
\phi_{l}^{+}(1)\phi_{m}^{+}(2)\phi_{k}^{+}(3)
{\bf A}_{12}{\bf A}_{13}
\phi_{l}(3)\phi_{m}(1)\phi_{k}(2)\; ,
\end{equation}
\begin{equation}
<H_{f}>=-\frac{e^{2}}{2c^{2}m}\sum_{l,m,k}\int
d1d2d3
|\phi_{l}(1)|^{2}\phi_{m}^{+}(2)\phi_{k}^{+}(3)
{\bf A}_{12}{\bf A}_{13}
\phi_{m}(3)\phi_{k}(2)
\end{equation}
where $d1=d^{2}{\bf r}_{1}$, sums extend over occupied Landau levels.
We assume that\\
$\sum_{l}|\phi_{kl}><\phi_{kl}| =\Pi_{k}$ is the k-th Landau level
projector.
Hartree-Fock self energies are found from the first variation of the
ground-state energy \cite{Fetter,Cabo}. Using methods developed for anyons one
obtains (in units of $E_{F}=\frac{2\pi\hbar^{2}\rho}{m}$ - the Fermi
energy of the 2D electron gas) \cite{Fetter,Sitko}:
\begin{displaymath}
H_{HF}=\frac{1}{n}\sum_{k=0}^{n-1}(k+\frac{1}{2}+2p+2p^{2}n-2p^{2}nE_{R}
+p^{2}nS_{n-1}-2p^{2}nS_{n-1-k})\Pi_{k}
\end{displaymath}
\begin{equation}
+\frac{1}{n}\sum_{k=n}^{\infty}(k+\frac{1}{2}-p^{2}+2p^{2}nE_{R}
+p^{2}n^{2}(\frac{1}{k}+\frac{1}{k+1})-p^{2}nS_{n-1}+2p^{2}nS_{k-n})
\Pi_{k}
\end{equation}
where $E_{R}=\ln{R\sqrt{\frac{eB^{ef}}{\hbar
c}}}+\frac{1}{2}(\gamma-\ln{2})$	($\gamma\simeq
0.577...$ - the Euler
constant), $R$ - a sample radius,
$S_{m}=\sum_{k=1}^{m}\frac{1}{k}$ and $S_{0}=0$.

Above results confirm the choice of the ground state. We find the
large energy gap ($\simeq 4p^{2}E_{R}$) between the last occupied
Landau level and the next ones. Taking $n=1$ we obtain the same results
as Cabo and Martinez  \cite{Cabo}. The ground-state energy
is given by:
\begin{equation}
<H_{a}>=\frac{N}{2},\;\;<H_{b}>=\frac{pN}{n},
\end{equation}
\begin{equation}
<H_{c}>+<H_{d}>=
2p^{2}NE_{R}+p^{2}NS_{n-1}-\frac{1}{2n}p^{2}N(n-1),
\end{equation}
\begin{equation}
<H_{e}>=p^{2}N\;\; ,
<H_{f}>=-2p^{2}NE_{R}-\frac{p^{2}}{n}(nS_{n-1}-n+1)N\;
{}.
\end{equation}
Hence, the total result is finite and equals:
\begin{equation}
\label{HFenergia}
<H>=\frac{1}{2}NE_{F}(1+3p^{2}+\frac{2p}{n}-\frac{p^{2}}{n}).
\end{equation}
In the limit $n\rightarrow\infty$ one finds the system in the zero
effective field
and the ground state is the Fermi sea \cite{Halperin}. The direct
calculations of the Hartree-Fock energy of this system \cite{Sitko2} give the
same
result as obtained from Eq.(\ref{HFenergia}), i.e. $<H>=
\frac{1}{2}NE_{F}(1+3p^{2})$ as $n\rightarrow\infty$.

The energy of electrons in the external magnetic field
$B^{ex}=\frac{2pn+1}{n} \frac{hc}{e}\rho$ is\\
 $\frac{1}{2}NE_{F}(2p+\frac{1}{n})$. Comparing with the result
(\ref{HFenergia})
one finds the difference:
\begin{equation}
\Delta
E=\frac{1}{2}NE_{F}[2p^{2}+(p-1)^{2}(1-\frac{1}{n})]
\end{equation}
which is always greater than zero. For example when $p=1$ the cost of
the generation of fluxes is $\Delta
E=NE_{F}$, i.e. twice the energy of the 2D electron gas (for $n=1$,
$\Delta E=NE_{F}p^{2}$).
Hence, the HFA suggests that
transmutations to composite fermions in
fractional-quantum-Hall-effect systems are not energetically
favourable at filling fractions $\nu=\frac{n}{2pn+1}$.

\noindent{\bf 3. RPA response function}

The Hamiltonian $H$ can be separated into two parts:
$H=H_{0}+H_{1}$
where\\ $H_{0}=\frac{1}{2m}\sum_{i}({\bf P}_{i}+{\bf A}^{ef}_{i})^{2}$ is
treated as the
unperturbed term and
$H_{1}$ is the {\it interaction} Hamiltonian	 \cite{Chen,Dai}.
Let us define current densities:
\begin{equation}
\bf J \rm ({\bf r})=\frac{1}{2m}\sum_{j}\left\{{\bf P} _{j}
+\frac{e}{c}{\bf A}_{j}+\frac{e}{c}{\bf A}_{j}^{ex},\delta({\bf r} -{\bf r}
_{j})\right\}
\end{equation}
and
\begin{equation}
\bf j \rm ({\bf r})=\frac{1}{2m}\sum_{j}\left\{{\bf P}_{j}
+\frac{e}{c}{\bf A}_{j}^{ef},\delta({\bf r} -{\bf r}
_{j})\right\}
\end{equation}
where braces denote an anticommutator.
 {\bf J}, {\bf j} are the vector parts
of $J^{\mu}$, $j^{\mu}$, respectively, with $\mu=0,x,y$. We define
$J^{0},j^{0}$ as density fluctuations:
\begin{equation}
j^{0}=J^{0}=\sum_{j}\delta
({\bf r} -{\bf r} _{j})-\rho.
\end{equation}
The problem of interest is the linear response of the system to a
weak external
electromagnetic field, $J^{\mu}=K^{\mu\nu}A_{\nu}$, which is characterized by:
\begin{equation}
K^{\mu\nu}({\bf q} ,\omega)=\frac{4\pi\rho e^{2}}{mc^{2}}
\delta^{\mu\nu}(1-\delta^{\mu 0})
+\frac{4\pi e^{2}}{\hbar c^{2}}\Delta_{R}^{\mu\nu}
({\bf q} ,\omega) .
\end{equation}
where	(in the real space)
\begin{equation}
\Delta_{R}^{\mu\nu}({\bf r} t,{\bf r'} t')=
-i\theta(t-t')<[J^{\mu}({\bf r} t),J^{\nu}({\bf r'} t')]> .
\end{equation}
Choosing ${\bf q}=q\hat{\bf x}$ and the Coulomb gauge one has
$J^{x}=\frac{\omega}{q}J^{0}$ and we can restrict our
problem to $2\times 2$ $K^{\mu\nu}$ matrix  \cite{Halperin}. It is
convenient first to consider the retarded correlation function of two
effective field currents:
\begin{equation}
D_{R}^{\mu\nu}({\bf r} t,{\bf r'} t')=
-i\theta(t-t')<[j^{\mu}({\bf r} t),j^{\nu}({\bf r'} t')]> .
\end{equation}
The random phase approximation consists in approaching $D_{R}$
by a sum of bubble graphs which can be written as:
\begin{equation}
D_{R}^{RPA}
({\bf q} ,\omega)=[I-D_{R}^{0}
({\bf q} ,\omega)V({\bf q})]^{-1}D_{R}^{0}
({\bf q} ,\omega)
\end{equation}
where $V$ is the interaction matrix  obtained from
the Hamiltonian $H_{1}$. We have	\\($\hbar$, $c$ $=1$,
$\omega_{c}=\frac{eB^{ef}}{m}$ )
\begin{equation}
V({\bf q} )=\frac{4p\pi}{q^{2}}\left(
\begin{array}{cc}
2pn\omega_{c}&-iq\\
iq&0
\end{array}\right) .
\end{equation}
As in the case of anyons  \cite{Jacak} one finds
\begin{equation}
D_{R}^{0}
({\bf q} ,\omega)=\frac{n}{2\pi\omega_{c}}\left(
\begin{array}{cc}
q^{2}\Sigma_{0}&-iq\omega_{c}\Sigma_{1}\\
iq\omega_{c}\Sigma_{1}&\omega_{c}\Sigma_{2}
\end{array}\right)
\end{equation}
where
\begin{displaymath}
\Sigma_{j}=\frac{e^{-x}}{n}\sum_{m=n}^{\infty}\sum_{l=0}^{n-1}
\frac{m-l}{(\frac{\omega}{\omega_{c}})^{2}-(m-l-i\eta)^{2}}
\frac{l!}{m!}x^{m-l-1}[L_{l}^{m-l}(x)]^{2-j}
\end{displaymath}
\begin{equation}
\times [(m-l-x)L_{l}^{m-l}(x)+2x\frac{dL_{l}^{m-l}(x)}{dx}]^{j}
\end{equation}
and $x=\frac{q^{2}}{2eB^{ef}}$.
Then we have:
\begin{equation}
D_{R}
({\bf q} ,\omega)=\frac{n}{2\pi\omega_{c}det}\left(
\begin{array}{cc}
q^{2}\Sigma_{0}&-iq\omega_{c}\Sigma_{s}\\
iq\omega_{c}\Sigma_{s}&\omega_{c}\Sigma_{p}
\end{array}\right)
\end{equation}
where
 $det=det(I-D^{0}V)=(1-2pn\Sigma_{1})^{2}-(2pn)^{2}\Sigma_{0}(1+\Sigma_{2})$,
 $\Sigma_{s}=\Sigma_{1}-2pn\Sigma_{1}^{2}+2pn\Sigma_{0}\Sigma_{2}$,\\
 $\Sigma_{p}=(2pn)^{2}\Sigma_{1}^{2}+\Sigma_{2}-(2pn)^{2}\Sigma_{0}
 \Sigma_{2}$.

Collective modes are determined by the zeros of the determinant $det$.
Because of the denominator in $\Sigma_{j}$ one finds an infinite
set of modes with gaps $\omega_{m}= m\omega_{c}$ as $q\rightarrow 0$,
$m=2,3, ...$
 \cite{Hanna2}, which is in agreement with results of Lopez and Fradkin
 \cite{Lopez2}. In contrast to the case of anyons we have no gapless
excitations.

The full current correlation function is given by \cite{Chen}:
\begin{equation}
\Delta_{R}\simeq\Delta_{R}^{RPA}=(I+U^{+})D_{R}^{RPA}(I+U)
\end{equation}
where
\begin{equation}
U=\frac{\omega_{c}2pn}{q}\left(\begin{array}{cc}
0&-i\\
0&0
\end{array}\right) .
\end{equation}
Hence, the linear response kernel equals:
\begin{equation}
K^{RPA}=\frac{e^{2}n}{2\pi det}\left(
\begin{array}{cc}
\frac{q^{2}}{\omega_{c}}\Sigma_{0}&-iq(\Sigma_{s}+2pn\Sigma_{0})\\
iq(\Sigma_{s}+2pn\Sigma_{0})&\omega_{c}(1+\Sigma_{2})
\end{array}\right) .
\end{equation}
As it is
expected there is no Meissner effect
in the system, i.e. $K^{yy}(\omega=0,q\rightarrow 0)=0$. Furthermore, the Hall
conductance of the
system is given by:
\begin{equation}
\sigma_{xy}(q,0)=
\frac{e^{2}n}{2\pi det}(-2pn\Sigma_{0}-\Sigma_{s})
\simeq|_{q\rightarrow
0}\;\frac{e^{2}}{2\pi}\;\frac{n}{1+2pn}=\frac{e^{2}}{2\pi}\nu
\end{equation}
which is a fractional multiple of $\frac{e^{2}}{h}$. Thus, the system
of composite fermions exhibits a fractional quantum Hall effect with
the correct value of
the Hall conductance. This result was first obtained by Lopez and
Fradkin \cite{Lopez1,Lopez2} within a field-theoretical approach.

\noindent{\bf 4. Conclusions}

Within the formalism developed for anyons we consider the
Chern-Simons theory of the FQHE. In the HFA we confirm the choice of
the ground state as proposed by Jain \cite{Jain,Greiter} within the mean
field approximation. We obtain the
large energy gap ($\simeq \ln{R}$) between the last occupied
Landau level and the next ones. The ground-state energy is found and in the
limit of $B^{ef}\rightarrow 0$ it agrees with the direct result for
the system in the zero effective field
 \cite{Sitko2}. Comparing with the energy of electrons in the
external magnetic field $B^{ex}=\frac{2pn+1}{n} \frac{hc}{e}\rho$ we
find transmutations to composite fermions in
fractional-quantum-Hall-effect systems not to be energetically
preferable at filling fractions $\nu=\frac{n}{2pn+1}$.

The RPA response function is found and gives the expected Hall
conductance in agreement with results of Lopez and Fradkin
 \cite{Lopez2}. We find also the set of collective modes with gaps
$m\omega_{c}$, however, a dependence on $q$ is not given.


\end{document}